\documentclass[aps, showkeys,showpacs, pra,twocolumn] {revtex4}
\usepackage{hyperref}
\usepackage{amsmath }
\usepackage{amssymb}
\usepackage{epsfig}
\usepackage{natbib}
\newcommand*{\be}{\begin{equation}}
\newcommand*{\ee}{\end{equation}}

\begin{document}
\bibliographystyle{revtex}

\title{Motion Caused by Magnetic Field in Lobachevsky Space}

\author{V.V. Kudryashov, Yu.A. Kurochkin, E.M. Ovsiyuk and V.M. Red'kov }
 \affiliation{B.I. Stepanov Institute of Physics,
National Academy of Sciences of Belarus,\\  68 Nezavisimosti Ave.,
220072 Minsk, Belarus}

\begin{abstract}
 We study   motion of  a relativistic particle  in the
3-dimensional Lobachevsky space  in the presence of an external
magnetic field which is  analogous to  a constant uniform magnetic
field in the Euclidean space. Three integrals of motion are found
and equations of motion are solved exactly in the special
cylindrical coordinates. Motion on surface  of the cylinder of
constant radius is considered in detail.
\end{abstract}



\maketitle

\section{INTRODUCTION}

The description of a particle behavior in a constant uniform
magnetic field is one of the basic problems in theoretical
physics.  As well known in the case of the Euclidean space, this
problem is exactly solved in the classical [1] and  quantum [2,3]
mechanics. Recently the generalization of the quantum mechanical
solution to the case of the Lobachevsky space was given [4,5].

 In the  present paper, an exact  description of a relativistic
 motion of a classical particle in an external magnetic field will
 be constructed on  the background
of the hyperbolic Lobachevsky $H_{3}$ space.
 This result  can be  used to describe the behavior of the charged
particles in macroscopic magnetic field in the context of
astrophysics.

\section{THE LAGRANGE EQUATIONS AND INTEGRALS OF MOTION}

We start with a metric of the Lobachevsky space in the special
cylindrical coordinates [6]:

\begin{eqnarray}
 dS^{2} = c^{2} dt^{2} -   \cosh^{2} (z / \rho) \; d r^{2}\nonumber \\
  -
\rho^{2} \cosh^{2} (z / \rho) \; \sinh^{2} (r / \rho) \;
 d \phi^{2}  - dz^{2} \; ,
 \nonumber
 \\
z \in ( - \infty , + \infty ), \quad r \in [0, +\infty ) , \quad
\phi \in [0, 2 \pi ] \; ,
 \label{1}
 \end{eqnarray}
 where $\rho$ is a curvature radius. The curvilinear coordinates
$(r, \phi,z)$ of the three-dimensional Lobachevsky space are
connected with quasi-Cartesian coordinates of the comprehending
four-dimensional pseudo-Euclidean space by relations

\begin{eqnarray}
 u_{1} = \rho \; \cosh(z / \rho)  \; \sinh(r / \rho)
\; \cos \phi \; , \nonumber \\
  u_{2} = \rho \; \cosh(z /
\rho) \; \sinh(r / \rho) \; \sin \phi \; , \nonumber \\
 u_{3} = \rho \;
\sinh(z / \rho) \; ,
\nonumber
\\
  u_{0} = \rho \; \cosh (z / \rho) \; \sinh
(r / \rho) \;,
\nonumber
\\
 u_{0}^{2} -
u_{1}^{2} - u_{2}^{2} - u_{3}^{3} = \rho^{2} \; , \quad u_{0} = +
\sqrt{\rho^{2}  + {\bf u}^{2} } \;  .
 \label{2}
\end{eqnarray}

The given metric determines the explicit expression for a squared
particle velocity

\begin{eqnarray}
\epsilon = \left({dz \over dt}\right)^{2} + \cosh^{2} (z / \rho )
\left({dr \over dt}\right)^{2} \nonumber \\
+ \cosh^{2} (z / \rho )\rho^{2} \sinh^{2} (r / \rho)
 \left({d \phi \over d t}\right)^{2}  .
\label{3}
\end{eqnarray}

In accordance with [4,5] we define a potential for a magnetic
field in the following form

\begin{equation}
A_{\phi} = - \rho^{2} B \; [ \; \cosh(r / \rho) -1 ]\; .
\label{4}
\end{equation}
This potential generalizes  a potential for a constant uniform
magnetic field with a strength $B$ in the Euclidean space when we
choose the direction of the field as the $z$ axis and use the
usual cylindrical coordinates. The potential (3) is the solution
of Maxwell's equations in the Lobachevsky space. Here these
equations are reduced to a single equation

\begin{eqnarray}
 {1 \over \sqrt{-g}}
{\partial \over \partial x^{r} }\sqrt{-g} \; F^{r \phi} = 0 \;,
\nonumber
\end{eqnarray}

\noindent
where

\begin{eqnarray}
 F_{\phi r} =
\partial_{\phi}A_{r} - \partial_{r} A_{\phi} \; ,
\nonumber
\\
 \sqrt{-g} =   \cosh^{2} (z / \rho) \; \sinh(r /  \rho ) \; .
\nonumber
\end{eqnarray}

The relativistic Lagrangian for a particle in above magnetic field
is of the form

\begin{eqnarray}
L = -mc^{2} \; \sqrt{1 - {\epsilon \over c^{2}}}\nonumber \\
 + \frac{e B \rho^{2}}{c}    [ \cosh(r / \rho) -1 ] \; \left({d \phi
\over d t}\right)\; . \label{5}
\end{eqnarray}

Then the  Lagrange equations of the second order are expressed as

\begin{eqnarray}
{d^{2} r \over dt^{2} } + 2  \; \tanh (z / \rho)  \; \left(dz
\over dt\right) \; \left(dr \over dt \right)
 = \rho \; \sinh (r / \rho)   \nonumber \\
\times  \left [ \; \cosh  (r / \rho) \; \left(d \phi \over
dt\right) + {\omega \over \cosh^{2} (z / \rho )} \; \right ]
\left(d \phi \over dt \right)  \; ,
\label{6}
\end{eqnarray}

\begin{eqnarray}
{d \over dt}\; \left [ \rho^{2}\;  \sinh^{2}(r / \rho ) \;
\cosh^{2} (z / \rho)\; \left(d \phi \over dt \right) \right.
 \nonumber \\
\left.  + \omega \rho^{2}\; (\; \cosh (r / \rho) -1 ) \right ] = 0
\; , \label{7}
\end{eqnarray}

\begin{eqnarray}
 {d^{2} z \over d t^{2}} -  {1 \over \rho } \; \cosh (z /
\rho) \; \sinh (z / \rho) \nonumber \\
\times \left  [\; \left({dr \over dt}\right)^{2} + \rho^{2} \;
\sinh^{2}(r /  \rho)\; \left({d \phi \over d t} \right)^{2} \;
\right ] =0 \; .
\label{8}
\end{eqnarray}
Here we have introduced notation for a quantity

\begin{equation}
\omega =
 {eB \over mc } \;  \sqrt{1 - {\epsilon \over c^{2}}} \; .
\label{9}
\end{equation}

The squared velocity $\epsilon$ is the integral of motion for the
considered system in the Lobachevsky space as the usual squared
velocity is the integral of motion in the Euclidean space.

It is obvious that a quantity

\begin{eqnarray}
I = \omega \rho^{2}\; [\; \cosh (r / \rho) -1 ] \nonumber
\\ + \rho^{2}\; \sinh^{2}(r / \rho ) \; \cosh^{2} (z / \rho)\; \left(d
\phi \over dt \right)   \label{10}
\end{eqnarray}
is conserved. This integral of motion is the generalization of the
usual angular momentum.

At last it is simply to verify that in addition to the conserved
quantities $\epsilon$  and   $I$, third quantity

\begin{equation}
A = \cosh^{4} (z /  \rho ) \left [  \left({dr \over dt}\right)^{2}
+ \rho^{2} \; \sinh^{2} (r /  \rho ) \left({d \phi \over
dt}\right)^{2}  \right ] \;
\label{11}
\end{equation}
is the integral of motion, at $\rho \rightarrow \infty$  it tends
to the squared velocity of motion perpendicular to $z$ axis.

\section{INTEGRATION OF THE EQUATIONS OF MOTION}

It should be stressed that the use of the special cylindrical
coordinates allows us to find the simple analytical expressions
for three integrals of motion. In the presence of these conserved
quantities it is easily to perform transition from the initial
equations of the second order (5) - (7) to the following equations
of the first order:

\begin{equation}
\left({dz \over dt}\right)^{2} = \epsilon - {A \over \cosh^{2} (z
/ \rho )} \; ,
\label{12}
\end{equation}

\begin{equation}
 \left({dr \over dt}\right) ^{2} = {A \over \cosh^{4} (z /
\rho )}\;  - { [  I - \omega \rho^{2}  (  \cosh (r / \rho ) -1 )
\;  ]^{2} \over \ cosh^{4} (z / \rho ) \rho^{2}
 \sinh^{2} (r / \rho )} \;   ,
\label{13}
\end{equation}

\begin{equation}
 {d \phi \over d t} = { I - \omega \rho^{2}\; (\;
\cosh (r / \rho ) -1 ) \over  \rho^{2} \;
 \sinh^{2}(r / \rho ) \;
\cosh^{2} (z / \rho ) } \; . \label{14}
\end{equation}

First of all we consider Eq. (12). The sign of the quantity
$\epsilon -A$ determines the character of solution.

If $\epsilon > A $, then  we have the explicit expressions for the
coordinate

\begin{equation}
\sinh (z(t) / \rho ) = \pm \; \sqrt{1 - {A \over \epsilon } }\;
\sinh ( \sqrt{\epsilon}  (t - t_{0})/ \rho )  \label{15}
\end{equation}
and the velocity

\begin{equation}
{d z(t) \over dt} = \pm \; \sqrt{ \epsilon - { A \epsilon \over
 (\epsilon - A) \; \sinh^{2} (\sqrt{\epsilon} (t- t_{0})/\rho) + \epsilon }}
 \;.
\label{16}
\end{equation}
In this case a particle passes through the point $z = 0$ with
velocity $dz /  dt = \pm \sqrt{\epsilon -A}$.

If $\epsilon < A $, then  we have other explicit expressions for
the coordinate

\begin{equation}
\sinh (z(t) / \rho ) = \pm \; \sqrt{{A \over \epsilon } -1 }\;
\cosh( \sqrt{\epsilon} (t - t_{0})/\rho) \label{17}
\end{equation}
and the velocity

\begin{equation}
{d z(t) \over dt} = \pm \; \sqrt{ \epsilon - { A \epsilon \over
 (A -\epsilon) \; \sinh^{2} (\sqrt{\epsilon} (t- t_{0})/ \rho) + A }} \; .
\label{18}
\end{equation}
In this case the particle moves either in the region

\begin{eqnarray}
\sinh (z / \rho ) >  \sinh (z_+ / \rho ) = \sqrt{A/ \epsilon -1}
\nonumber
\end{eqnarray}
or in the region
\begin{eqnarray}
\sinh (z / \rho ) <  \sinh (z_- / \rho ) = - \sqrt{A/ \epsilon
-1}.
\nonumber
\end{eqnarray}
In other words there is the repulsion at the points $z_-$ and
$z_+$ leading to existence of the forbidden region for motion.

It should be noted that  velocity along $z$ axis is constant in
the case of the Euclidean space ($\rho \rightarrow \infty$)
 whereas  velocity along $z$ axis is variable in the case of the
Lobachevsky space.

From the equations (12) - (14) it is seen that after obtaining the
explicit dependence $z(t)$, the further solution of the problem is
reduced to the following integrations

\begin{eqnarray}
\int { \rho  \sinh (r / \rho ) \; dr \over \sqrt{w(r)}}\nonumber \\
 =
 \pm \; \int {dz \over  \cosh(z / \rho )\; \sqrt{ \epsilon  \cosh^{2} (z / \rho ) - A }} \; ,
\label{19}
\end{eqnarray}

\begin{equation}
\phi - \phi_{0} =  \pm  \int {  [ I - \omega \rho^{2} ( \cosh (r /
\rho ) -1  )  ] \; d r  \over \rho \; \sinh  (r / \rho )
\sqrt{w(r) }} \; , \label{20}
\end{equation}
where

\begin{eqnarray}
 w(r)=A \rho^{2} \sinh^{2} (r / \rho ) - [ I -
\omega \rho^{2} \;
 (  \cosh(r / \rho ) - 1  )  ] ^{2}.
\nonumber
\end{eqnarray}

 Both formula (19) for the dependence $r(z)$  and formula
(20) for the dependence $\phi (r)$   give the particle orbit.

Now we examine the connection of $\phi$ and $r$ in detail. The
integration in (20) leads to the following relation

\begin{eqnarray}
(I + \omega  \rho^{2})  \cosh (r / \rho )  - \omega \;\rho^2
\nonumber \\
= \sqrt{C} \;\rho \sinh (r / \rho )  \cos (\phi - \phi_{0}) \; ,
\label{21}
\end{eqnarray}
where

\begin{equation}
C = A - \rho^{2} \omega^{2} + \frac{1}{\rho^{2}}(I  + \omega
\rho^2)^{2} \; . \label{22}
\end{equation}

A quantity $C$ as function of $A$ and $I$ is the integral of
motion and has the form

\begin{eqnarray}
C =  \cosh^{4} (z /  \rho )\;  \left({dr \over dt}\right)^{2} +
\rho^{2} \; \sinh^{2} (r / \rho) \nonumber \\
\times \left [\; \cosh (r / \rho ) \;\; \cosh^{2} (z /  \rho) \;
\left(d \phi \over dt\right) + \omega \; \right ]^{2} \; ,
\label{23}
\end{eqnarray}
$$
C \geq 0 \; .
$$

\section{THE  MOTION ON THE CYLINDRICAL SURFACE}

Let us compare formula (21) with the well known formula

\begin{eqnarray}
\cosh (R / \rho ) \; \cosh (r / \rho ) - \cosh (r_{0} / \rho )
\nonumber \\
= \sinh (R / \rho )\; \sinh(r / \rho) \; \cos (\phi - \phi_{0})
\label{24}
\end{eqnarray}
describing in the Lobachevsky space  a circle of radius $r_{0}$,
if a distance between a circle center and the origin of
coordinates is $R$. Formulas (21) and (24) coincide, if the
parameters $R$ and $r_0$ are connected with the integrals of
motion $A$ and $I$ by the relations

\begin{equation}
\cosh (r_{0} / \rho ) = {1 \over  \sqrt{1 - A / \omega^{2}
\rho^{2} }} \; ,
\label{25}
\end{equation}
\begin{equation}
 \cosh (R / \rho ) = { I + \omega \rho^{2} \over  \omega
\rho^{2}   \sqrt{1 - A / \omega^{2} \rho^{2}}} \; . \label{26}
\end{equation}

We emphasize that the expressions (25) and (26) become sensible,
if the inequality

\begin{equation}
 A < \omega^{2} \rho^{2}
\label{27}
\end{equation}
is accomplished. Then in the $(r, \phi)$ - projection, a particle
moves along the circle. The circle center is shifted with respect
to the origin. Taking into account a motion in the $z$ direction
we get a complete  motion on the cylindrical surface.
 The  coincidence of a coordinate origin and a circle center
 corresponds to the zeroth value of the integral of motion $C$. In
 this case a $z$ axis coincides with a  cylinder axis.

 The transition to this coordinate system leads to the considerable
 simplification. From the condition $C=0$  we see that

\begin{equation}
{dr \over dt } = 0 \; , \quad (r = r_{0}) \; ,
\label{28}
\end{equation}

\begin{equation}
  {d \phi \over d t} = - { \omega \over  \cosh (r_{0} / \rho)\; \cosh^{2} (z / \rho ) } \; .
 \label{29}
 \end{equation}

In this case the integrals of motion $A$ and $I$ are expressed via
the circle radius $r_0$  as follows

 \begin{equation}
A = \omega^{2} \rho^{2} \; \tanh^{2} (r_{0} / \rho )\; ,
\label{30}
\end{equation}
\begin{equation}
  I = \omega \rho^{2} \; \left[ \; {1 \over
\cosh (r_{0} / \rho )}  -1 \right] \; . \label{31}
\end{equation}
Taking into account the relations (15) - (18) we get the explicit
formulas  describing evolution of an angle  and an angular
velocity.

If $\epsilon > A$, we obtain the following expressions

 \begin{eqnarray}
 \phi(t) - \phi_{0} =   {- \omega \rho \over
\cosh (r_{0} / \rho ) \; \sqrt{A} } \nonumber \\
\times  \mbox{artanh}\; \left [ \; \sqrt{{A \over \epsilon } } \;
\tanh ( \sqrt{\epsilon} (t - t_{0} )/ \rho ) \; \right ] \; ,
\label{32}
\end{eqnarray}

\begin{equation}
{d \phi (t) \over dt } =  { - \omega \epsilon \over  \cosh (r_{0}
/ \rho )  [(\epsilon - A)  \sinh^{2} (\sqrt{\epsilon}(t - t_{0})/
\rho ) + \epsilon ]}
\\[2mm]
\label{33}
\end{equation}
and if $\epsilon < A$, we obtain other expressions

 \begin{eqnarray}
 \phi(t) - \phi_{0} =   {- \omega \rho \over \cosh \;(r_{0} /
\rho ) \; \sqrt{A} }
\nonumber
\\
 \times \mbox{artanh} \; \left [ \; \sqrt{{\epsilon
\over  A } } \; \tanh ( \sqrt{\epsilon}  (t - t_{0} )/ \rho ) \;
\right ] \; ,
\label{34}
\end{eqnarray}

\begin{equation}
{d \phi (t) \over dt } =  { - \omega \epsilon \over  \cosh (r_{0}
/ \rho ) [(A- \epsilon ) \; \sinh^{2} (\sqrt{\epsilon} (t -
t_{0})/ \rho ) + A  ]} \; .\label{35}
\end{equation}

In the case of the Euclidean space the angular velocity is
constant whereas in the case of the Lobachevsky space the angular
velocity is variable and tends to zero at $t \rightarrow \infty$.

\section{Conclusion}

In conclusion several additional  remarks should be given. It can
be shown that the formula (21)  for trajectories in the form $F(r,
\phi)=0$ describes not only the motions on cylindrical surfaces,
shifted or not with respect to the initial axis $z$, but also
describes  motions infinite in radial variable $r$. So, there
exist two  different classes of trajectories, finite and infinite
in radial variable.

Besides, it can be  demonstrated that the potential of the
magnetic field (4)  is invariant
 with respect to shifts in  the Lobachevsky
space in the plane perpendicular to the axis $z$. This symmetry
underlies the structure of two families  of  possible solutions of
the equations of motion,  providing us with a transitivity group
acting
 within two classes of motions.

 All results can be extended to a spherical  Riemann model $S_{3}$
 on the base of the metric
 \begin{eqnarray}
 dS^{2} = c^{2} dt^{2} -   \cos^{2} (z / \rho) \; d r^{2}\nonumber \\
  -
\rho^{2} \cos^{2} (z / \rho) \; \sin^{2} (r / \rho) \;
 d \phi^{2}  - dz^{2} \; ,
 \nonumber
 \\
z \in [ - \pi / 2, + \pi / 2  ] , \quad r \in [0, +\pi ] , \quad
\phi \in [0, 2 \pi ] \
 \label{36}
 \end{eqnarray}
and a potential for a magnetic field   in $S_{3}$ according to

\begin{equation}
A_{\phi} =  \rho^{2} B \; [ \; \cos(r / \rho) -1 ]\; .
\label{37}
\end{equation}

\section{Acknowledgments}

  The authors are grateful to E.~A. Tolkachev for valuable comments and useful
  discussion. This work has been supported by the Belarusian
  Republican Foundation for Fundamental Research (Project No.
  F09D - 009).

\end{document}